\documentclass[longauth]{aa}  

\usepackage{ltxtable}
\usepackage{longtable}
\usepackage{graphicx}
\usepackage{rotating}
\usepackage{pdflscape}
\usepackage{color}
\usepackage{natbib}
\usepackage{amssymb}
\usepackage{lscape}
\usepackage{txfonts}
\begin{document}

   \title{The {\it Gaia}-ESO Survey: Probes of the inner disk abundance gradient
  \thanks{Based on data products from observations made with ESO
    Telescopes at the La Silla Paranal Observatory under programme ID
    188.B-3002 and 193.B-0936. These data products have been processed by the Cambridge Astronomy Survey Unit (CASU) at the Institute of Astronomy, University of Cambridge, and by the FLAMES/UVES reduction team at INAF/Osservatorio Astrofisico di Arcetri. These data have been obtained from the {\it Gaia}-ESO Survey Data Archive, prepared and hosted by the Wide Field Astronomy Unit, Institute for Astronomy, University of Edinburgh, which is funded by the UK Science and Technology Facilities Council.}$^{,}$\thanks{Full Table
    2 is only available in electronic form
at the CDS via anonymous ftp to cdsarc.u-strasbg.fr (130.79.128.5)
or via http://cdsweb.u-strasbg.fr/cgi-bin/qcat?J/A+A/}}

   \titlerunning{Inner disk metallicity gradient}

   \author{H.~R. Jacobson\inst{1}, E.~D. Friel\inst{2}, 
     L. J\'{i}lkov\'{a}\inst{3},
     L. Magrini\inst{4}, A. Bragaglia\inst{5}, A. Vallenari\inst{6},
     M. Tosi\inst{5}, S. Randich\inst{4}, P. Donati\inst{5,7},
     T. Cantat-Gaudin\inst{6,8}, R. Sordo\inst{6},
     R. Smiljanic\inst{9}, J.~C. Overbeek\inst{2}, 
     G. Carraro\inst{10},
     G. Tautvai\v{s}ien\.{e}\inst{11}, 
     I. San Roman\inst{12}, S. Villanova\inst{13},
     D. Geisler\inst{13}, C. Mu\~{n}oz\inst{13}, 
     F. Jim\'{e}nez-Esteban\inst{14,15},
     B. Tang\inst{13}, 
     G. Gilmore\inst{16},
     E.~J. Alfaro\inst{17},
     T. Bensby\inst{18},
     E. Flaccomio\inst{19},
     S.~E. Koposov\inst{16,26}
     A.~J. Korn\inst{20},
     E. Pancino\inst{4,21},
     A. Recio-Blanco\inst{22},
     A.~R. Casey\inst{16},
     M.~T. Costado\inst{17},
     E. Franciosini\inst{4},
     U. Heiter\inst{20},
     V. Hill\inst{22},
     A. Hourihane\inst{16},
     C. Lardo\inst{23},
     P. de Laverny\inst{22},
     J. Lewis\inst{16},
     L. Monaco\inst{24},
     L. Morbidelli\inst{4},
     G.~G. Sacco\inst{4},
     S.~G. Sousa\inst{25},
     C.~C. Worley\inst{16},
     S. Zaggia\inst{6}
     }

   \institute{
   Massachusetts Institute of Technology, Kavli Institute of Astrophysics \& Space Research, Cambridge, MA, USA\\
   \email{hrj@mit.edu}
   \and
   Department of Astronomy, Indiana University, Bloomington, IN, USA
   \and
   Leiden Observatory, P.O. Box 9513, NL-2300 RA Leiden, The Netherlands
   \and
   INAF--Osservatorio Astrofisico di Arcetri, Largo E. Fermi, 5, I-50125, Florence, Italy
   \and
   INAF-- Osservatorio Astronomico di Bologna, Via Ranzani, 1, 40127, Bologna, Italy
   \and 
   INAF- Osservatorio Astronomico di Padova, Vicolo Osservatorio 5, 35122 Padova, Italy
   \and
   Dipartimento di Fisica e Astronomia,  Via Ranzani, 1, 40127, Bologna, Italy
  \and
   Dipartimento di Fisica e Astronomia, Universita di  Padova, vicolo Osservatorio 3, 35122 Padova, Italy
\and
   Department for Astrophysics, Nicolaus Copernicus Astronomical Center, ul. Rabia\'nska 8, 87-100 Toru\'n, Poland 
\and
   European Southern Observatory, Alonso de Cordova 3107 Vitacura, Santiago de Chile, Chile
   \and
     Institute of Theoretical Physics and Astronomy, Vilnius University, A. Gostauto 12, 01108 Vilnius, Lithuania
   \and
   Centro de Estudios de F\'{i}sica del Cosmos de Arag\'{o}n (CEFCA),
   Plaza San Juan 1, E-44001 Teruel, Spain
   \and
   Departamento de Astronom\'{i}a, Casilla 160-C, Universidad de
   Concepci\'{o}n, Concepci\'{o}n, Chile
   \and
   Centro de Astrobiolog\'{i}a (INTA-CSIC), Dpto. de Astrof\'{i}sica,
   PO Box 78, 28691 Villanueva  de la Ca\~{n}ada, Madrid, Spain
   \and
   Suffolk University, Madrid Campus, C/ Valle de la Vi\~{n}a 3, 28003
   Madrid, Spain
   \and
   Institute of Astronomy, University of Cambridge, Madingley Road,
   Cambridge CB3 0HA, United Kingdom
   \and
   Instituto de Astrof\'{i}sica de Andaluc\'{i}a-CSIC, Apdo. 3004,
   18080, Granada, Spain
   \and
   Lund Observatory, Department of Astronomy and Theoretical Physics,
   Box 43, SE-221 00 Lund, Sweden,
   \and
   INAF - Osservatorio Astronomico di Palermo, Piazza del Parlamento 1, 90134, Palermo, Italy
   \and
    Department of Physics and Astronomy, Uppsala University, Box 516,
    SE-751 20 Uppsala, Sweden,
   \and
   ASI Science Data Center, Via del Politecnico SNC, 00133 Roma,
   Italy,
   \and
   Laboratoire Lagrange (UMR7293), Universit\'e de Nice Sophia
   Antipolis, CNRS,Observatoire de la C\^ote d'Azur, CS 34229,F-06304
   Nice cedex 4, France,
   \and
   Astrophysics Research Institute, Liverpool John Moores University,
   146 Brownlow Hill, Liverpool L3 5RF, United Kingdom,
   \and
   Departamento de Ciencias Fisicas, Universidad Andres Bello,
   Republica 220, Santiago, Chile,
   \and
   Instituto de Astrof\'isica e Ci\^encias do Espa\c{c}o, Universidade
   do Porto, CAUP, Rua das Estrelas, 4150-762 Porto, Portugal
   \and
   Moscow MV Lomonosov State University, Sternberg Astronomical Institute, Moscow 119992, Russia}
   \authorrunning{Jacobson et al.}
   \date{}

 
  \abstract
   {The nature of the metallicity gradient inside the solar circle
     ($R_{\rm GC} < 8$ kpc) is poorly understood, but studies of
     Cepheids and a small sample of open clusters suggest that it steepens
   in the inner disk.}
   {We investigate the metallicity gradient of the inner disk using a
     sample of inner disk open clusters that is three times larger than has previously
     been studied in the literature to better characterize the
     gradient in this part of the disk.}
   {We used the {\it Gaia}-ESO Survey (GES) $\rm[Fe/H]$ values and stellar
     parameters for stars in 12 open clusters in the inner 
    disk from GES-UVES data.  Cluster mean $[\rm Fe/H]$ values were determined
   based on a membership analysis for each cluster.  Where necessary,
   distances and ages to clusters were determined via comparison
   to theoretical isochrones.}
   {The GES open clusters exhibit a radial metallicity gradient of
     $-0.10\pm0.02$ dex kpc$^{-1}$, consistent with the gradient measured by
     other literature studies of field red giant stars and open
     clusters in the range $R_{\rm GC}\sim 6-12$
     kpc. We also measure a trend of increasing $[\rm Fe/H]$ with increasing
     cluster age, as has also been found in the literature.}
   {We find no evidence for a steepening of the inner disk metallicity
   gradient inside the solar circle as earlier studies indicated.
   The age-metallicity relation shown by the clusters is consistent
   with that predicted by chemical evolution models that include the
   effects of radial migration, but a more detailed comparison between
   cluster observations and models would be premature.}


   \maketitle
%

\section{Introduction}\label{intro}

Ever since the seminal work of \citet{janes1979}, open clusters have
been used to trace the metallicity distribution in the Galactic disk.
In that work, Janes found that the distribution of metallicity ($\rm [Fe/H]$) with
Galactocentric distance ($R_{\rm GC}$) was consistent with a gradient with
metallicity decreasing with increasing distance from the Galactic
center at a rate of $-$0.05 dex kpc$^{-1}$.  Such a distribution has
been shown in chemical evolution models to represent the inside-out
growth of the Galactic disk (e.g., \citealt{chiappini_model97}).
The term ``gradient'' has come to be used as short hand to refer
to the (azimuthally averaged) change of $[\rm Fe/H]$
 as a function of $R_{\rm GC}$ and has generally been
 described as a linear function, as in \citet{janes1979}. However, it
 has been pointed out several times over the decades that the
 metallicity distribution with $R_{\rm GC}$ shown by open clusters and other
 stellar populations can be described by other functions, as well (e.g.,
 \citealt{taat97}, \citealt{And02inner}).  Indeed, as the number of
 clusters studied by high-resolution spectroscopy has grown over the
 past decade, it now seems clear that the form of the radial
 dependence on mean metallicity in the disk changes with
 Galactocentric radius.  While a linearly decreasing function may well
 describe the change in $\rm [Fe/H]$ as a function of $R_{\rm GC}$ through
 the solar neighborhood out to $R_{\rm GC} \sim 10-13$ kpc,  the
 clusters in the outer disk have roughly constant $\rm [Fe/H] \sim -0.3$ and remarkably small dispersion (e.g., \citealt{Heiter2014}).

Much work has been done in recent years to better characterize the
gradient in the outer disk, and the nature of the transition
between the inner and outer disk (e.g., \citealt{carraro2007},
\citealt{yong2012}, \citealt{brag2008}, \citealt{jacobson2011}, \citealt{cantat-gaudin2016}).  It can be said, however, that the gradient in the inner disk has received comparatively less attention.  This is not surprising given that the line of sight toward the Galactic center is much more crowded, confused, and extincted.  Nevertheless, this is an important part of the Galaxy where the bulge/bar meets both the thin and thick disks.

There is some evidence that the nature of the gradient changes inside
the solar circle.  Some of the first evidence came from Cepheids:
\citet{And02inner} found that the gradient inside $R_{\rm GC} \sim 7-8$ kpc
was steeper than that outside the solar circle.  Recent work by
\citet{genovali2013,genovali2014} has reinforced the increase in
      $\rm [Fe/H]$ seen in the inner disk to super solar metallicities of
      $\sim 0.4$ dex.  The Cepheid sample of \citet{martin15}, which
      extends to as close as 3 kpc from the Galactic center, indicates
      that 0.4 dex is the upper bound of the metallicity range in the
      inner disk, and the gradient plateaus at this level inside of
      $R_{\rm GC} = 5$ kpc.
      In contrast to the evidence from Cepheids, a study of inner disk
      red clump stars has found no evidence for a gradient at all: red
      clump stars spanning $R_{\rm GC} \sim 4.5-6.5$ kpc along a line of
      sight towards the Galactic center have a mean $\rm [Fe/H] = +0.17$
      with no sign of a change with $R_{\rm GC}$\ \citep{Hill2012inner}.  

Regarding the open cluster population, the number of confirmed
clusters inside $R_{\rm GC} \sim 7$ kpc that we could use to probe the inner disk gradient is small relative to that in the solar neighborhood and beyond.  Many works have noted the dearth of old ($>$1 Gyr) clusters in the inner region  (see, e.g., \citealt{Bonatto06}) and while new surveys in the infrared are revealing promising candidates for future study (e.g., \citealt{Froebrich2007}, \citealt{Borissova2014}), many of these candidate clusters are embedded and highly obscured.  Relatively few have fundamental parameters determined, and of those even fewer are accessible to traditional optical techniques for abundance analysis.  However, there have been efforts to study the known open clusters in the inner disk specifically  to determine whether the older population of open clusters shows the same steepening of the gradient that is seen in the younger Cepheids.
With their sample of three clusters all inside $R_{\rm GC} \sim 7$ kpc, along with a selection of clusters from the literature, \citet{magrini2010}  did indeed find evidence that the gradient rises more sharply towards the galactic center.  While this work greatly increased the number of inner disk open clusters subject to detailed chemical abundance study, the inner disk cluster sample that has been well studied is still small relative to that in the solar neighborhood and beyond.

In the last few years,  
a number of spectroscopic surveys have been planned and begun, 
aimed at investigating and
constraining in great detail the chemo-dynamical evolution of the
Milky Way: for example, the extentions to SDSS
\citep{york_sdss} SEGUE \citep{segueI} and APOGEE {\citep{apogee};
RAVE \citep{rave}; the {\it Gaia}-ESO Survey \citep{gilmore2012,Randich2013};
GALAH \citep{galah}.  Several multi-object spectrographs
have been designed and/or built that will also carry out their own
spectroscopic surveys:
 LAMOST \citep{zhao_lamost,cui_lamost}; 
WEAVE \citep{weave1,weave2}; MOONS \citep{moons}; 4MOST \citep{4most}.  
In particular,
the SDSS III APOGEE Survey \citep{ahn2014,apogee} has done much to increase our
  picture of the metallicity distribution of the disk, based
  on near-infrared high-resolution spectroscopy of red giant stars.
  While the APOGEE Survey includes many open clusters, their sample
  predominantly has objects located at $R_{\rm GC} \gtrsim 7$ kpc \citep{2013_apogeegrad,
    cunha2016}, and in some cases only a few members per cluster were observed.  Instead, insight into the inner disk metallicity
  distribution comes from a study of APOGEE field stars, which found that the
gradient of $\sim -0.09$ dex kpc$^{-1}$ in the Solar neighborhood
transitions around $R_{\rm GC} = 6$ kpc,  becoming {\it shallower} interior to that value \citep{hayden2014,hayden2015}.

The {\it Gaia}-ESO Survey (GES)
is a large-scale, high-resolution spectroscopic survey being carried out with
the FLAMES instrument on the VLT.  In addition to probing field stars in all major stellar components of the Milky Way, the survey includes a large number of open clusters spanning a wide range in age and
location in the Galactic disk (\citealt{gilmore2012},
\citealt{Randich2013}).   
Stars of all evolutionary states are observed in GES, from
  pre-main sequence stars to red giants, and elements from all groups
  of the Periodic Table, from Li to the neutron-capture species, are
measurable in the FLAMES spectra.
Regarding open clusters, with the complementary capabilities of the
GIRAFFE and UVES spectographs, GES stellar
  samples can range from 10$^{2}$ to 10$^{3}$ members per cluster
  (e.g., \citealt{donati2014,cantatM11,frasca2015,prisinzano2016}) allowing not only for robust average measures of cluster
  chemistry and dynamics, but also their variations as a function of stellar
  evolutionary state \citep{smiljanic2016}.

Though the survey is far from complete, the
first internal data releases  from GES (data releases iDR1, iDR2/3,
iDR4) span roughly 2.5 out of five years of observations and provide
an opportunity to explore the form of the abundance gradient
in the inner disk from high-resolution spectroscopy of both open
clusters and field stars.   Complementary studies of the thin and
thick disk gradients and age-metallicity relationships using GES iDR1 data
for field stars have appeared in \citet{bergemann2014},
\citet{Recio2014} and \citet{mikolaitis14}.
\citet{bergemann2014} used UVES data of field FGK stars to explore the
age-metallicity relationships and gradients in the thin and thick disk
populations.  To obtain ages, they limited their sample to unevolved
stars and so probed only the solar neighborhood at distances of $7 < 
R_{\rm GC} < 9$ kpc.   \citet{Recio2014} used the lower resolution GIRAFFE
spectra of FGK stars in the first data release to study the kinematic
and spatial gradients of the thin and thick disk populations over the
range  $5 < R_{\rm GC} < 12$ kpc. \citet{mikolaitis14} used a subset of
the \citet{Recio2014} sample and the distances calculated therein
that spanned a range of $4 < R_{\rm GC} < 12$ kpc and found a rather
shallow metallicity gradient for the thin disk ($\sim -0.03$ to $-0.04$ dex kpc$^{-1}$).

While current sample sizes of field stars in spectroscopic
  surveys such as GES and 
  APOGEE reveal much about the state of our Galaxy,
open clusters within these surveys are complementary tracers of the
chemistry of the thin disk because their ages and distances
can be precisely determined, they span a wide range in age and
location within the Galaxy, and their chemical abundances can be robustly
measured from large samples of stellar members.
In this work, we utilize the iDR4 GES results obtained from UVES
observations of stars in 12 intermediate-aged ($>100$ Myr, $<3$ Gyr)
open clusters.  These 12 are from a total of 26 open clusters included in iDR4.
These results are placed
in the context of those of other inner disk populations to explore the
nature of the abundance gradient in this relatively understudied region of the
Galaxy.  This paper is organized as follows: Sect.~\ref{ges}
presents the GES observations used in this work; Sect.~\ref{grad}
presents the discussion of the inner disk metallicity gradient and
Sect.~\ref{mod_disc} compares it to
chemical evolution models.
  A summary and conclusions are available in Sect.~\ref{summary}.

\section{GES data}\label{ges}

The {\it Gaia}-ESO survey began on 31 December 2011.  
The survey makes use of the FLAMES instrument with both GIRAFFE and
the fiber link to UVES.  Here, we used only the UVES spectra obtained
with the 580 nm setup, which
have spectral resolution $\rm R \equiv \lambda/\Delta\lambda$ =
47\,000, signal-to-noise (S/N) $\gtrsim50$ and span 4760--6840 \AA.
UVES data were processed with the GES pipeline
as described in \citet{SaccoGES}.  Stellar parameters and abundances for FGK-type stars observed with UVES 
were determined by up to 14 independent groups, and homogenized in a
process described in \citet{ges_uves}.  Results from  working groups specializing in different stellar types were then combined to result in a set of
recommended parameters utilized by the GES collaboration (Hourihane et
al. in prep).  These recommended parameters, available in the 
``recommendedastroanalysis'' table within GESiDR4Final, were used in this
work.

Analysis has now been completed and
the data products for all survey observations taken before July 2014 have been
released to the consortium as GESviDR4Final.  This data release
included some updated analysis methods and homogenization across multiple
working groups.  As a result, this release included newly
determined stellar parameters and abundances for the previously
analyzed objects, so the parameters presented here differ slightly from
those in earlier GES publications.

Twelve intermediate-aged clusters inside the solar circle (for this
work we adopted $R_{\rm GC} = 8.0$ kpc for the Sun) have been
targeted with UVES observations and included in GESviDR4Final (Table
1).  
Of these, results from earlier internal data releases were published
for the clusters Trumpler 20, NGC 4185, NGC 6705 (M 11) and Berkeley
81.  References for these works are given in Table~\ref{tab_params}.
For these objects, we adopted the ages and distances determined
from those published papers.  For five of the remaining eight clusters we 
used values from the literature as given in Table~\ref{tab_params}.  Three of these clusters are nearby and 
have distances that rest on Hipparcos measurements (NGC 2516, NGC 3532, and NGC 6633).  Two others (NGC 6005 and Pismis 18) have parameters that rely on traditional methods of fitting isochrones, but with models that are 
consistent with the metallicities derived here; we explore the impact of errors on their distances below.  For the remaining three clusters, we
estimated new cluster ages, distances, and $E(B-V)$ values taking into 
account the new GES metallicities, which are significantly higher than 
previously assumed metallicities (Be 44 and NGC 6802), or for which we already had 
analysis underway (Trumpler 23, Overbeek et al. in preparation).   Analysis for these clusters was intended only to give preliminary values that primarily take into account
the revised metallicity and membership information from GES; a more complete 
treatment and redetermination of cluster parameters will be presented
in future publications (e.g., Tang et al. in preparation for NGC 6802). 
For these clusters, we used radial velocities measured from both UVES 
and GIRAFFE observations in the cluster field to determine estimates of cluster systemic 
velocities.  First, a mean radial velocity was calculated from all
stars in the cluster field.
Probable radial velocity members were chosen as those stars within approximately 2 sigma 
about the mean velocity.  
Using this indication of membership in conjuction with published photometry
(\citealt{JanesHoq} for Be 44 and NGC 6802 and \citealt{carraro2006} for Trumpler 23), we fit PARSEC isochrones \citep{PARSEC} at the derived GES metallicities (see below), 
to obtain estimates of cluster reddening, distance, and age which are summarized in Table~\ref{tab_params}.  

For clusters available in earlier GES
releases, we determined membership independently of those
previously published.  Not only have the recommended parameters and
abundances varied slightly with each data release as a consequence of
the calibration process, but in the case of some clusters, additional fields were observed and include more cluster members than were available in iDR1 and iDR2/3.
Mean cluster metallicities were based on an evaluation of individual
star cluster membership using radial velocities and abundances from
UVES spectra in GESviDR4Final.      We used the radial velocities in
combination with the derived $\rm [Fe/H]$ values from UVES
observations as a guide for determining membership; cluster members
clump tightly in a diagram of radial velocity versus [Fe/H], aiding
the elimination of potential non-member stars near the cluster
velocity.   The majority of $\rm [Fe/H]$ values came from evolved stars in the red clump or along the giant branch, but for the closest clusters NGC 2516, 3532 and 6633, include determinations from dwarf stars as well.  The resulting mean cluster abundances and standard deviations about the mean and radial velocities,  based on these likely members, are given in Table~\ref{tab_params}.  The GES ID, coordinates, and the GES recommended parameters for all stars considered cluster members and used in forming the mean metallicities are given in Table~\ref{member_params}.

\begin{table*}
\caption{Cluster parameters}             
\label{tab_params}      
\setlength{\tabcolsep}{0.02in}
\centering                          
\scriptsize
\begin{tabular}{l l l l l l l l l l}        
\hline\hline                 
Cluster & \it{l} & \it{b} & Age & $\rm [Fe/H]$ &RV & \# stars &  $R_{\rm GC}$ & $z$ & Ref.\\    
         & ($^{\circ}$) & ($^{\circ}$)  &  (Gyr) & (dex) &  (km s$^{-1}$) &  &  (kpc) & (pc) &  \\
\hline                        
Berkeley 44 & 53.2 & $+$3.33 & 1.6$\pm$0.3 & $+$0.17$\pm$0.04 & $-$8.7$\pm$0.7 & 4 & 6.91$\pm$0.12 & 128$\pm$17 & This Study \\
Berkeley 81 & 34.51 & $-$2.07 & 0.86$\pm$0.10 & $+$0.21$\pm$0.06 & $+$48.3$\pm$0.6 & 13 & 5.49$\pm$0.10 & $-$126$\pm$7 & 1,2 \\
NGC 2516 & 273.8 & $-$15.8 & 0.16$\pm$0.04 & $-$0.06$\pm$0.05 & $+$23.6$\pm$1.0 & 15 & 7.98$\pm$0.01 & $-$97$\pm$4 & 3 \\
NGC 3532 & 289.6 & 1.35 & 0.30$\pm$0.10 & $-$0.03$\pm$0.02 & $+$4.8$\pm$1.4 & 2& 7.85$\pm$0.01 & 12$\pm$1 & 4  \\
NGC 4815 & 303.63 & $-2.10$ & 0.57$\pm$0.07 & $-$0.03$\pm$0.06 & $-$29.6$\pm$0.5 & 5 & 6.94$\pm$0.04 & $-$95$\pm$6 & 5  \\
NGC 6005 & 325.8 & $-$3.00 & 1.20$\pm$0.30 & $+$0.16$\pm$0.02 & $-$24.1$\pm$1.34 & 12 & 5.97$\pm$0.34 & $-$141$\pm$26 &  6  \\
NGC 6633 & 36.0 & 8.3 & 0.63$\pm$0.10 & $-$0.05$\pm$0.06 & $-$28.8$\pm$1.5 & 11 & 7.71$\pm$0.01 & 52$\pm$2 &  7 \\
NGC 6705 & 27.31 & $-2.78$ & 0.30$\pm$0.05 & $+$0.08$\pm$0.05 & $+$34.9$\pm$1.6 & 27 & 6.33$\pm$0.16 & $-$95$\pm$10 &  8 \\
NGC 6802 & 55.3 & 0.92 & 1.0$\pm$0.1 & $+$0.10$\pm$0.02 & $+$11.9$\pm$0.9 & 8 & 6.96$\pm$0.07 & 36$\pm$3 & This Study \\
Pismis 18 & 308.2 & 0.30 & 1.2$\pm$0.4 & $+$0.11$\pm$0.02 & $-$27.5$\pm$0.7 & 6 & 6.85$\pm$0.17 & 12$\pm$2 & 6 \\
Trumpler 20 & 301.48 & $+2.22$ & 1.50$\pm$0.15 & $+$0.10$\pm$0.05 & $-$40.2$\pm$1.3 & 42 & 6.86$\pm$0.01 & 136$\pm$4 & 9 \\
Trumpler 23 & 328.8 & $-$0.50 & 0.8$\pm$0.1 & $+$0.14$\pm$0.03 & $-$61.3$\pm$0.9 & 10 & 6.25$\pm$0.15 & $-$18$\pm$2 & This Study \\
\hline                                   
\end{tabular}
\tablefoot{References for cluster ages, $R_{\rm GC}$\ and $z$ distances: (1) \citet{DonatiBe81}; (2) \citet{magrini2015}; (3) \citet{sung2516}; (4) \citet{clem3532}; (5) \citet{friel2014}; (6) \citet{piatti98}; (7) \citet{jeffries6633}; (8) \citet{cantatM11}; (9) \citet{donati2014}}
\end{table*}

The location of these inner disk GES clusters relative to the plane of
the Milky Way disk is shown in the top panel of Fig.~\ref{z_rgc}.  All clusters are
within $\sim150$ pc of the plane, and appear to be thin disk objects.
The bottom panel of Fig.~\ref{z_rgc} shows the range of cluster age
as a function of $R_{\rm GC}$.  Clusters at similar $R_{\rm GC}$ can vary in age by
as much as 1 Gyr.

\begin{figure}
\centering
\includegraphics[width=1.0\hsize]{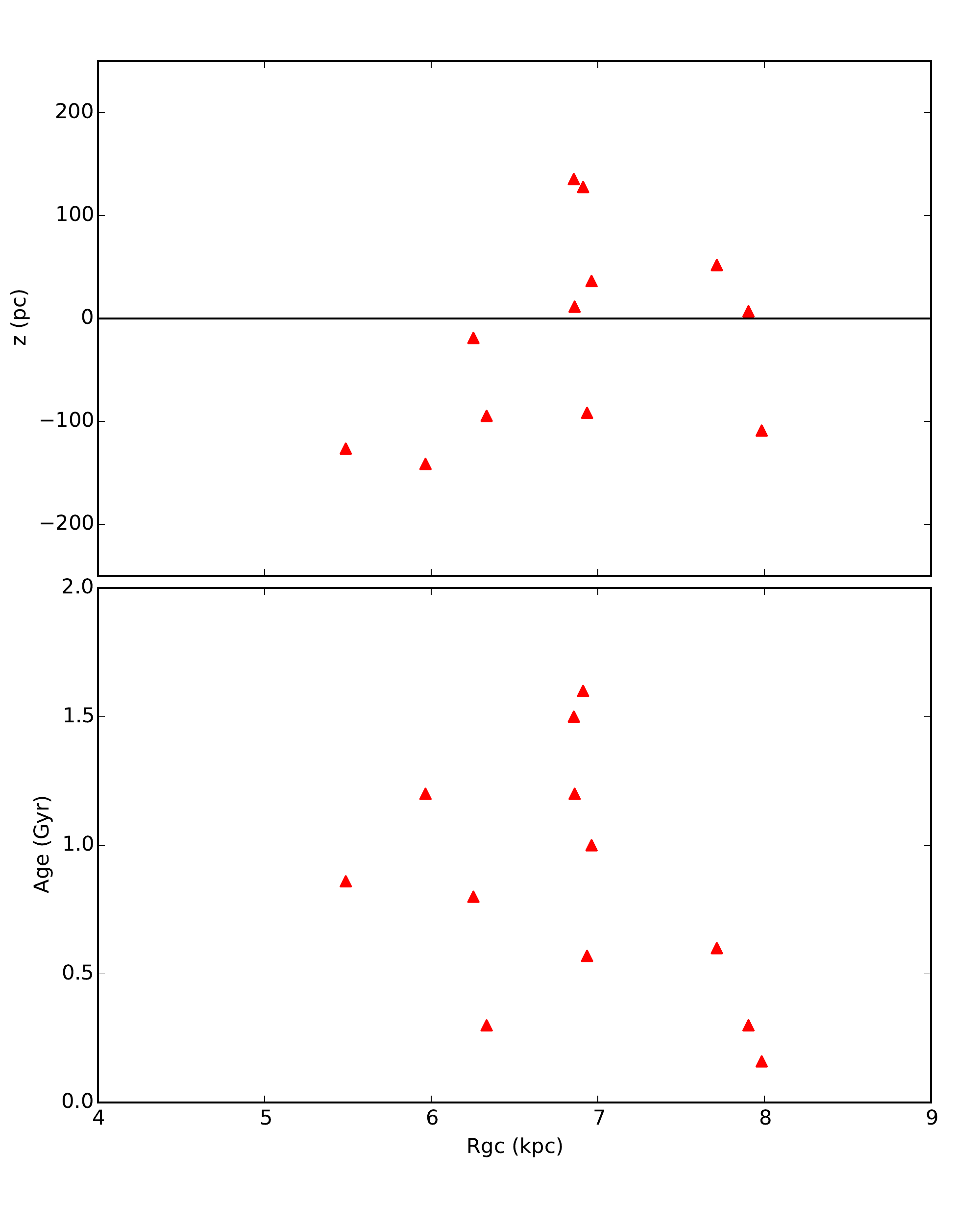}
\caption{Top panel: locations of the GES open clusters (red triangles)
  relative to the Galactic plane ($z=0$, solid line),
  as a function of their Galactocentric radii.  All clusters lie
  within 150 pc of the mid-plane, and thus are members of the thin
  disk. Bottom panel: cluster ages (in Gyr) plotted as a function of
  $R_{\rm GC}$.}
\label{z_rgc}
\end{figure}

\section{Inner disk metallicity gradient}\label{grad}

With nine open clusters inside $R_{\rm GC}=7$ kpc, the GES sample is a
factor of three times larger than previous homogeneous samples in this
part of the Galactic disk \citep{magrini2010} and allows us to see more 
clearly the trend of metallicity inside the solar 
radius.   As shown in Fig.~\ref{oc_grad}, the sample of 12 GES clusters (in red triangles)
shows a steady decrease in $\rm [Fe/H]$ with increasing Galactocentric distance over 
the range of $R_{\rm GC} \sim 5.5-8$ kpc.    The vertical error bars represent the 
standard deviations about the mean abundance determined from those stars judged 
to be members, and given in Table 1.   As discussed in
\citet{magrini2014} and \citet{magrini2015}, in light of the highly
uniform abundance scale for the GES determinations and the internal
chemical homogeneity of the clusters, the differences in metallicity
seen among the GES clusters are reliable and significant.  The current
sample reinforces this fact; as can be seen from the stellar
parameters given in Table~\ref{member_params}, typical uncertainties
in $\rm [Fe/H]$ for individual stars are 0.10 dex.  The rms deviations about the cluster means, on the other hand, are much smaller, from 0.02 to 0.06 dex, indicating that there is no evidence of an internal spread in abundance within each cluster.  There is a clear decrease in mean abundance from the innermost point sampled, at $R_{\rm GC}=5.5$ kpc, out to the clusters close to the solar neighborhood.  A linear regression analysis on the GES sample shows a metallicity
gradient with a slope of $-0.10 \pm$ 0.02 dex kpc$^{-1}$.   

In addition to this clear trend, however, there is significant
variation in $[\rm Fe/H]$ around $R_{\rm GC}=7$ kpc.  The  5 clusters at
$R_{\rm GC}=7$ kpc cover a range of 0.2 dex, from Be 44 at $\rm [Fe/H] = +0.17
\pm 0.04$ to NGC 4815, with $\rm [Fe/H]=-0.03 \pm 0.06$, well in excess
of the scatter about the mean cluster metallicities.  Since stars in
these two clusters have similar atmospheric parameters and were
analyzed homogeneously, this dispersion in $[\rm Fe/H]$ is likely real.  Of
the five clusters around  $R_{\rm GC}=7$ kpc, two have {\it l}
$\sim$ 54$^{\circ}$, and the other three cluster around {\it l}
$\sim$ 305$^{\circ}$ (Table 1).  However these two groups do not also
cluster together in metallicity: the $[\rm Fe/H]$ dispersion
at each Galactic longitude is 0.07 and 0.13 dex, respectively. 

As mentioned previously, this homogeneous GES cluster sample explores
the inner part of the disk with considerably larger statistics
than in previous studies.  More clusters will
be available in future GES releases, sampling also the outer parts of
the disk. 
At this stage, in order to study the behavior of the metallicity
distribution across a wider range of Galactocentric distances, 
we also consider the large compilation of high-resolution
open cluster metallicities in \citet{netopil2016}.  This compilation
is an update to that of \citet{Heiter2014}, who determined weighted
average $[\rm Fe/H]$ values for clusters based on rigorous selection criteria
over a limited range of atmospheric parameters of literature
measures.  Though the numbers of open clusters observed homogeneously
as part of large spectroscopic surveys (e.g., in GES; APOGEE -
\citealt{2013_apogeegrad}; LAMOST - \citealt{zhang2015}) 
is growing, the \citet{Heiter2014} sample and its update is
currently the closest one can come to a uniform sample formed from
inhomogeneous literature measurements. 
Eight clusters in the current work are also in
\citet{netopil2016}; we use the $[\rm Fe/H]$, age and $R_{\rm GC}$ values given in
Table~\ref{tab_params} for the clusters in common.

Figure~\ref{oc_grad} shows the metallicity distribution with $R_{\rm GC}$ for both
the GES inner disk clusters (red triangles), and the
\citet{netopil2016} sample (open gray squares).  We note
that though clusters beyond $R_{\rm GC} = 14$ kpc have been studied, we limit
our discussion to objects inside this limit, as the
gradient appears to change around this Galactocentric radius (e.g., 
\citealt{yong2005,yong2012,carraro2007,pancino2010,andreuzzi11,donati2015,netopil2016}; see also \citealt{taat97}).
For comparison, we
also plot open clusters observed as part of the APOGEE Survey
\citep{2013_apogeegrad} (filled gray squares)\footnote{We show only
  those APOGEE clusters in which more than one star was observed.}.

The linear gradient traced by the GES clusters, $-0.10\pm0.02$ dex kpc$^{-1}$  is shown as the dashed red line.  The gradient described
by the APOGEE cluster sample is just as steep, if not steeper;
\citet{2013_apogeegrad} reported $-0.20 \pm 0.08$ dex kpc$^{-1}$ for clusters
inside $R_{\rm GC} = 10$ kpc\footnote{\citet{2013_apogeegrad} adopted
$R_{\rm GC} = 8.5$ kpc for the Sun, as opposed to 8.0 kpc in this work.}.  
However, these cluster samples overlap in a very limited region of
$R_{\rm GC} \sim 7-8$ kpc, 
and given the small sample sizes these slopes may not fully reflect
the metallicity distribution in the inner disk.   Fitting their
larger literature sample \citet{netopil2016} find a linear metallicity
gradient of  $-0.085 \pm 0.017$ dex kpc$^{-1}$  for the 64 clusters
inside 9 kpc, a slope quite consistent with the relationship traced by
the GES clusters.  They note the strong influence of the three
clusters with much higher metallicity at $R_{\rm GC} \sim 7$ kpc (NGC 6253, 6583, and 6791)
and omitting those find a more modest slope of  $-0.061 \pm 0.015$ dex kpc$^{-1}$.  

We note that, while there is overlap in the distribution of metallicities of the various samples, the GES abundances appear on the lower boundary of  the distribution of  \citet{netopil2016}  and  APOGEE values at the same $R_{\rm GC}$.  The slightly lower $[\rm Fe/H]$ values for the GES
clusters around $R_{\rm GC} \sim8$ kpc may indicate a 
$\sim$0.1 dex zero-point offset
between the GES and \citet{netopil2016} metallicity scales.  
It is also possible that the small numbers of the GES clusters in
  the solar neighborhood shown here do not fully reflect its intrinsic abundance
  distribution, and a larger sample of clusters would include objects
  of solar metallicity.
For the eight clusters in common between the GES and the
\citet{netopil2016} study, the difference in $\rm [Fe/H]$ is only $0.04 \pm 0.05$ (s.d.), suggesting that the systematic differences are small.    Given these
caveats, the metallicity distribution of the GES clusters is
consistent with that of the \citet{netopil2016} sample.

\begin{figure} 
\centering
\includegraphics[width=1.0\hsize]{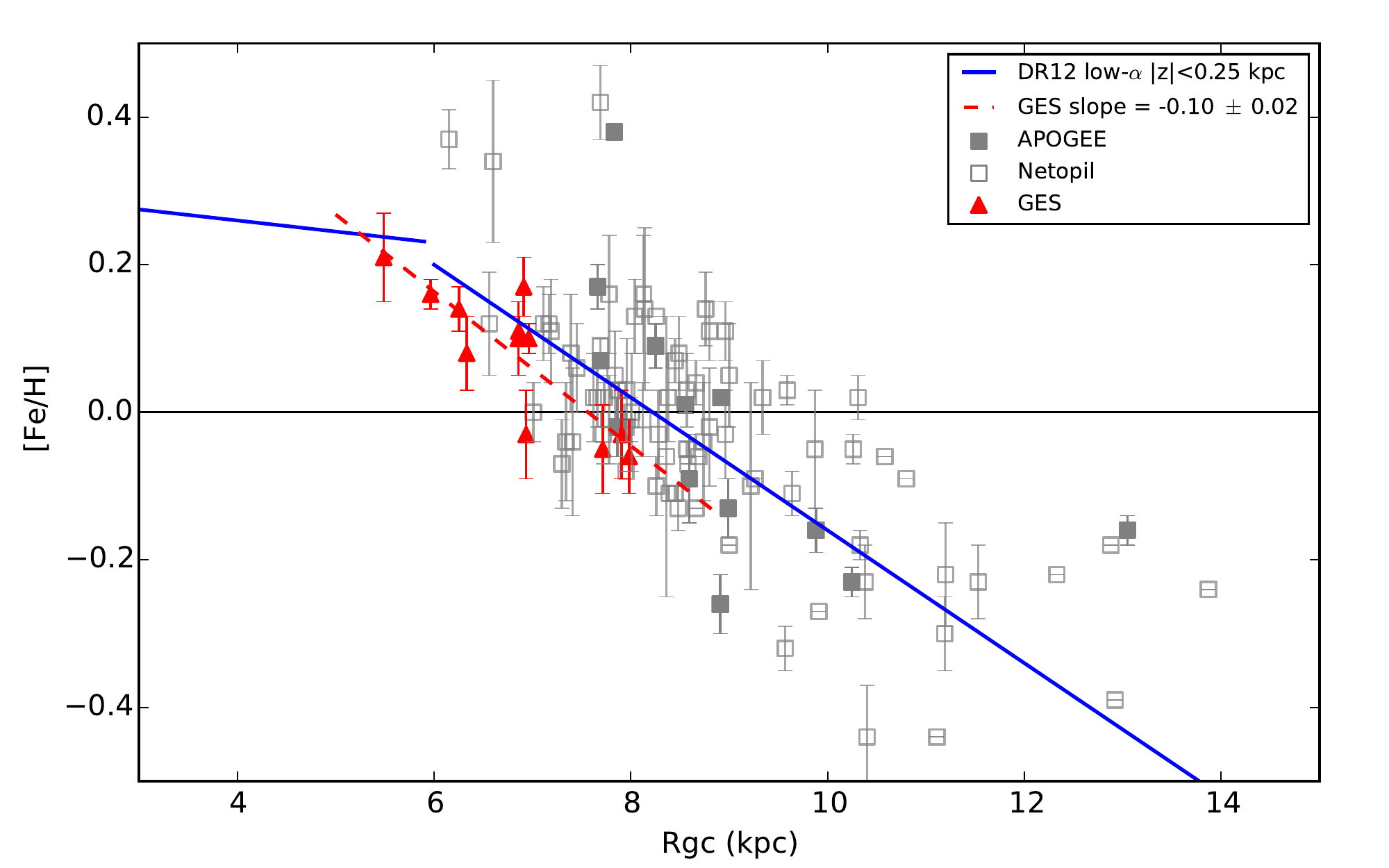}
\caption{Metallicity gradient as shown by the GES open clusters
  (red triangles)in
  comparison to literature studies.  The open cluster literature
  compilation of \citet{netopil2016} is shown as open gray squares,
  while clusters observed by the APOGEE survey \citep{2013_apogeegrad}
  are filled gray squares.  The solid blue line shows the metallicity
  gradient as determined by APOGEE field giant stars within 250 pc of
the Galactic mid-plane \citep{hayden2014}.  The red dashed line is
least-squares fit to the GES sample, with the value of the slope given
in the caption. $[\rm Fe/H] = 0$ is indicated by the solid black line, to
guide the eye.}
\label{oc_grad}
\end{figure}

Also shown in Fig.~\ref{oc_grad} is the gradient determined from
$2 \times 10^{4}$ giant stars in the APOGEE Survey
\citep{hayden2014}.  Specifically, we plot the gradient determined
from a subsample of APOGEE stars with low [$\alpha$/Fe] measures,
confined to within 250 pc of the Galactic mid-plane (i.e., likely thin
disk giant stars).  As mentioned in the Introduction, \citet{hayden2014}
found that the metallicity gradient of $-$0.09 dex kpc$^{-1}$ broke
and became shallower inside $R_{\rm GC} = 6$ kpc.  Only one GES object lies
inside (though on the edge of) this break, and its $[\rm Fe/H]$ value is
consistent with both slopes shown by the APOGEE data.   

\citet{hayden2014} also traced the vertical abundance gradients in the disk, 
finding that they may be steeper in the inner Galaxy, with values of $-$0.4 dex kpc$^{-1}$ 
for the low [$\alpha$/Fe] stars inside $R_{\rm GC} = 7$ kpc.   Figure~\ref{z_rgc} shows that the 
GES clusters reach at most 150 pc from the Galactic plane, so the impact of this gradient, 
if it applied to the cluster population, would be at most $\sim0.05$ dex.  Correcting for
this small effect, though, would serve to steepen the gradient by a small amount, 
to $-$0.12 dex kpc$^{-1}$ since the two innermost clusters are also among those most distant from the Galactic plane.  

While the importance of homogeneous $[\rm Fe/H]$ measures to the determination
of the metallicity gradient cannot be overstated, distances must also
be homogeneous.  Given that the GES clusters and the
\citet{netopil2016} sample have different distance scales (though both
adopt $R_{\rm GC} = 8.0$ kpc for the Sun), we
calculate the GES cluster metallicity gradient again adopting
\citet{netopil2016} $R_{\rm GC}$ values for objects in common.  This is shown
in Fig.~\ref{oc_grad2}.  The gradient steepens slightly to $-0.11\pm0.02$ dex
kpc$^{-1}$, as the \citet{netopil2016} $R_{\rm GC}$ values are generally
closer to the Sun than our values.  We noted earlier that two of the clusters in the GES sample 
have larger than typical uncertainties in distance; these have little impact on the derived linear gradient. 
Changing their distances by the quoted errors changes the gradient by at most 0.01.  We conclude that, in general, the metallicity
gradients shown by the GES sample, the \citet{netopil2016} sample, and
the APOGEE field star sample are consistent with one another.

\begin{figure}
\centering
\includegraphics[width=1.0\hsize]{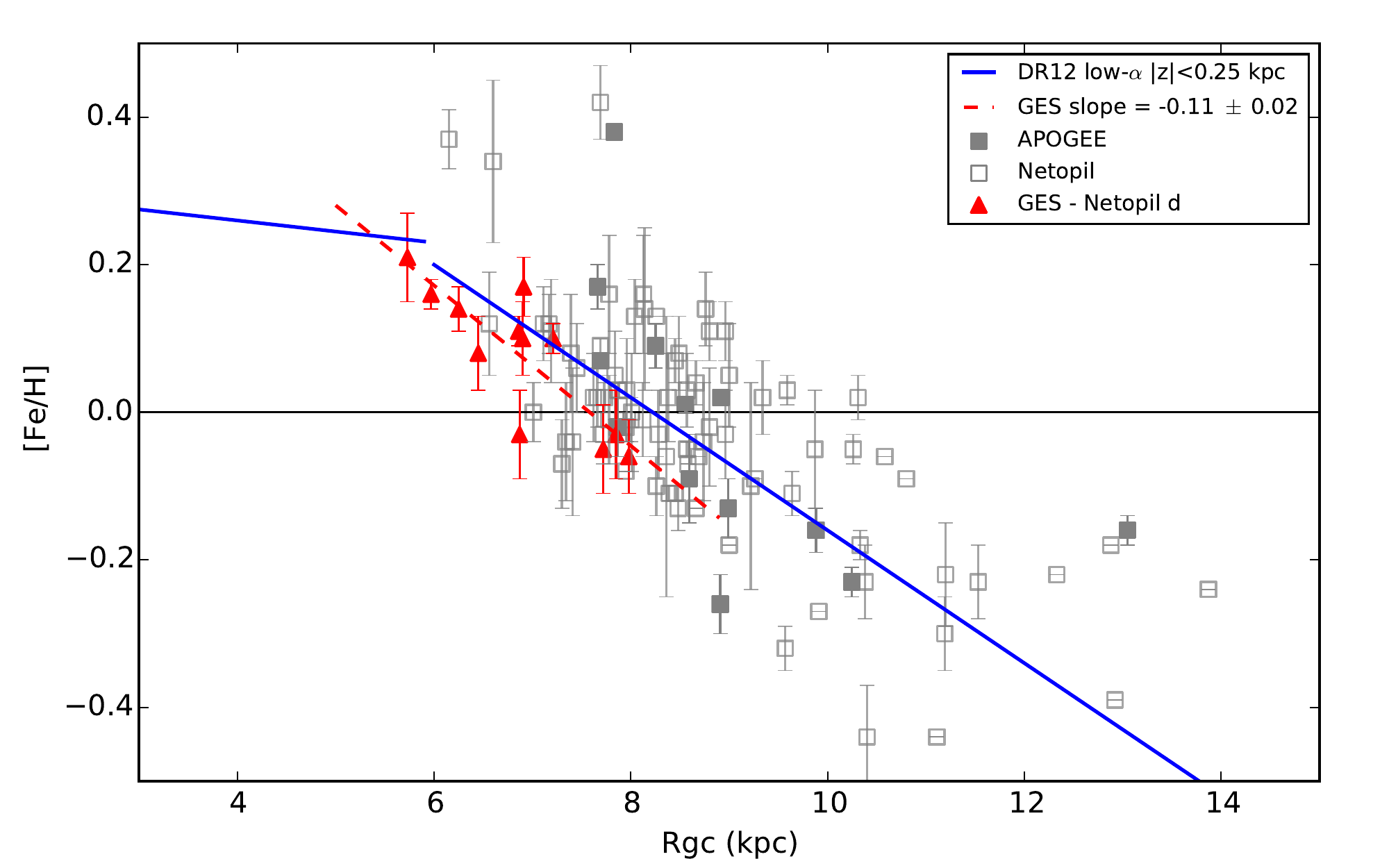}
\caption{Same as in Fig.~\ref{oc_grad}, but with distances determined by
  \citet{netopil2016} for GES clusters in their compilation.  Note the
magnitude of the GES cluster gradient has changed by 0.01 dex
kpc$^{-1}$.}
\label{oc_grad2}
\end{figure}

The motivation of this work was to determine if the GES inner disk
objects showed an indication of the metallicity gradient steepening
inside the solar circle as previous studies have indicated (e.g.,
\citealt{genovali2014,magrini2010}).  Given the consistency of the
slopes of the different samples in Fig.~\ref{oc_grad} and
Fig.~\ref{oc_grad2}, there is no indication that the open clusters' metallicity
gradient changes within the range of $R_{\rm GC} \sim 6-8$ kpc, and in fact
the APOGEE field star data indicate a shallower slope in the innermost
region studied.

\section{Comparison to model predictions}\label{mod_disc}

Open clusters have long been used to help constrain chemical evolution
models  of the
Milky Way disk,  through both the shape of their radial gradients and
their evolution with time (e.g., \citealt{tosi1982,Chiappini2001,cescutti2007,magrini2009}).  
To date, the observational evidence for the time evolution of the
gradient has been mixed.  Generally, open cluster studies have found
relatively little evidence of a change of the gradient with time,
though there are some indications that the younger clusters have a
slightly shallower distribution than the older clusters (a flattening
of the gradient with time: \citealt{friel2002}, \citealt{magrini2009},
\citealt{cp11,andreuzzi11}).  However, studies of other objects or combinations of different stellar populations have indicated the opposite.  \cite{sh2010}, for example, found evidence that the gradient has grown steeper with time based on an analysis of disk planetary nebulae. 

The chemical evolution model of 
\citet{minchev2013,minchev2014} allows for the movement of star
particles from their birthplace due to radial migration effects.
(We take their work as an example; other dynamical models of Milky
Way-like galaxies that trace
stellar chemistry do exist, e.g., 
\citealt{roskar2008,sb09,loebman2011,bird2012,kubryk2013}.)
Figure~\ref{grad_mod} shows the metallicity gradients predicted by their
model for different-aged objects within 250 pc of the Galactic plane (taken from Fig.~10 and Table 1 of
\citealt{minchev2014}).  We also plot the GES and \citet{netopil2016}
open cluster samples color-coded with the same age ranges as the model
gradients.  The majority of open clusters are $<2$ Gyr old (magenta
symbols), and show a steeper gradient than the model at all age
ranges.  We note also that the open clusters, almost regardless of age, do not align with the 
model predictions, either in the value of the metallicity or the
dependance on age.  It has been shown that open cluster metallicity does not correlate with
cluster age as would arise from a general increase in the 
metallicity of the disk with time (e.g., \citealt{friel2002}) and this general impression is reinforced in Fig.~\ref{grad_mod};
 almost all the younger clusters, with ages  $<2$ Gyr,
have lower metallicities than predicted in the model.  Conversely, the oldest of the clusters lie at 
higher metallicities than predicted.  

\citet{netopil2016}, investigating possible age-metallicity
relationships among their homogenized literature sample, found
evidence, in fact, of an increase in metallicity with age.  Comparing
metallicities of clusters with ages less than 0.5 Gyr to those with
ages between 1 and 2.5 Gyr, corrected for the radial gradient, they
found that the younger group had metallicities $\sim 0.07$ dex lower
than the older group.  This held even when they only considered
clusters within $7 \leq R_{\rm GC} \leq 9$ kpc.
We can carry out the same exercise with the GES
clusters, and find an identical result.  Residuals from the linear
gradient show a clear correlation with cluster age
(Fig.~\ref{grad_resids}); the younger clusters are at systematically
lower metallicities, while the older ones are at systematically higher
metallicities.  Using the same age groupings adopted by
\citet{netopil2016}, we find, as they did, that clusters with
ages less than 0.5 Gyr show a mean residual of $-0.021 \pm 0.026$
while clusters older than 1.0 Gyr have a mean residual of $+0.041 \pm
0.037$.  That this trend is evident in our homogeneously analyzed 
(and smaller) cluster sample that extends to $R_{\rm GC} \sim5.5$ kpc is
worth emphasizing.

As \citet{netopil2016} point out, this increase in metallicity with age,
contrary to what would be expected from simple expectations of
chemical evolution, can be explained by the effect of radial
migration, as more metal-rich and older objects from the inner disk
move outward.  An analogous result has been seen in a comparison of metallicity
gradients traced by planetary nebulae and H\,II regions in M 31: the
older nebulae exhibit a flatter gradient and higher oxygen abundances
than the younger H\,II regions \citep{magrini2016}.  In their
comparison to the chemical evolution model of \citet{minchev2013},
\citet{netopil2016} found that the trend of increasing metallicity
with age was qualitatively consistent with the effect of radial
migration, but was stronger than predicted by the model.  The same
holds true for our cluster sample.

\begin{figure}
\centering
\includegraphics[width=1.0\hsize]{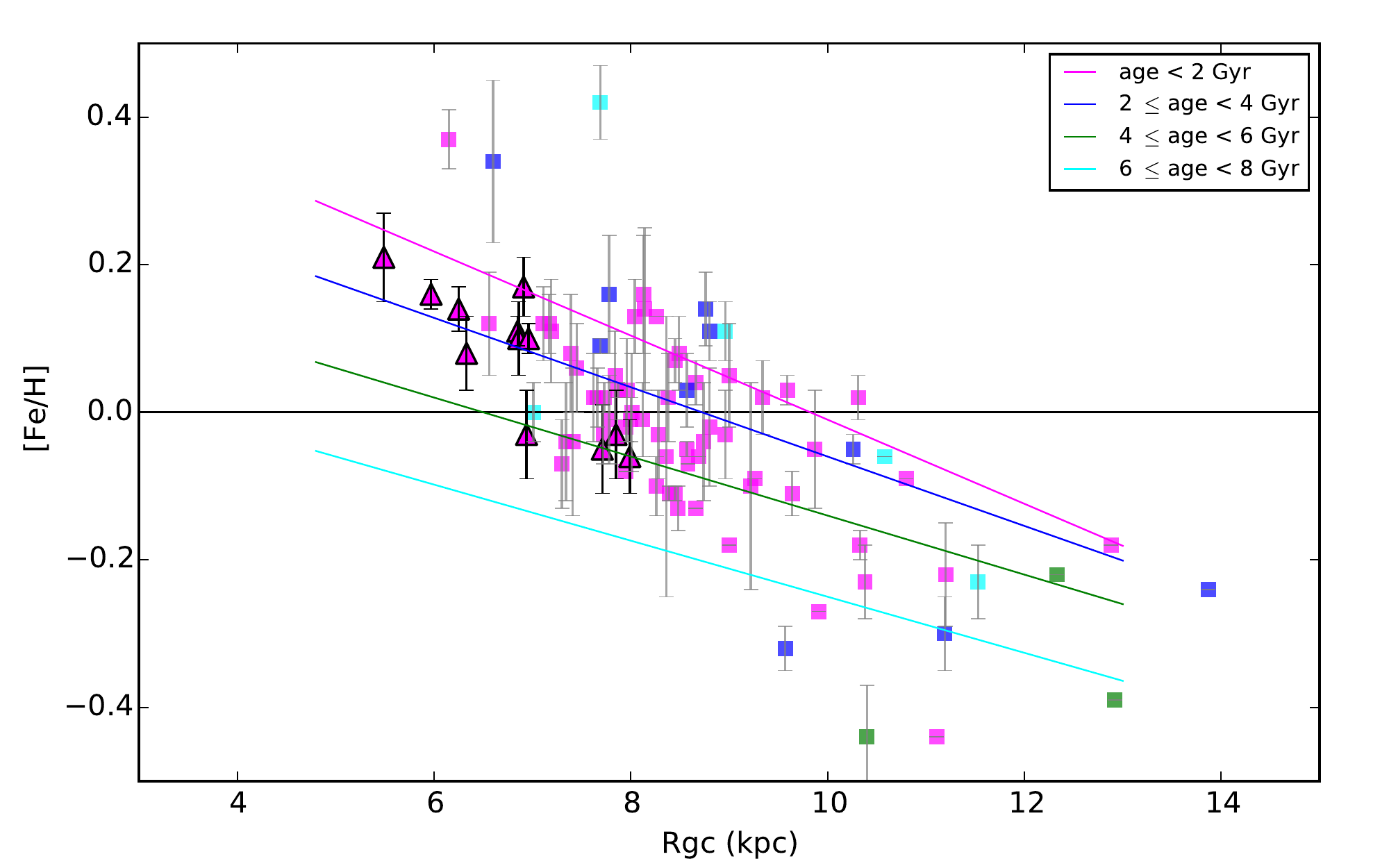}
\caption{GES open clusters (black-lined triangles) and
  \citet{netopil2016} clusters (filled squares), color-coded by age.
  The colored solid lines represent the metallicity gradients for
  different-aged stars in the chemical evolution model of
  \citet{minchev2014}.  The gradient shown by the clusters is much
  steeper than that of the model.}  
\label{grad_mod}
\end{figure}

\begin{figure}
\centering
\includegraphics[width=1.0\hsize]{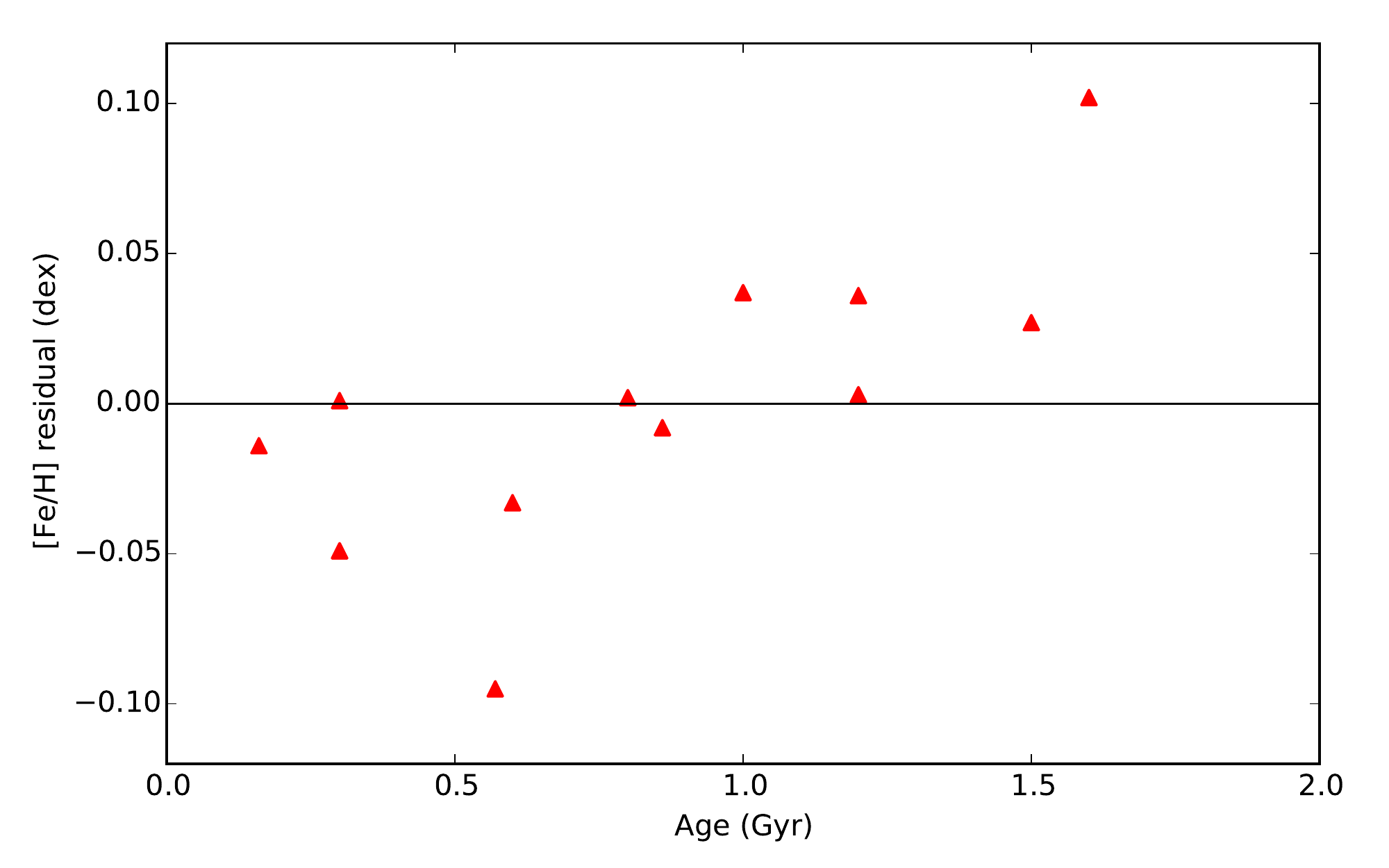}
\caption{Residuals in $\rm [Fe/H]$ from the linear gradient as a function of cluster age. }
\label{grad_resids}
\end{figure}

However, in the larger context, the exact magnitude and effect of
radial migration on the chemical abundance trends seen in the Milky
Way have yet to be firmly established.  Observationally, evidence
appears to be growing that radial migration effects cannot be
ignored.  
The lack of an age-metallicity relation for stars in
  the solar neighborhood can be naturally explained by migration
  processes, for example (e.g., \citealt{haywood2006}).
Also, the shapes of
metallicity distribution functions in different radial bins of APOGEE
red giant stars led  
\citet{hayden2015} to conclude that ``migration is of global importance
in the evolution of the disk'' (p.~12).
Furthermore, some radial migration  models can reproduce
observations remarkably well (e.g., \citealt{loebman2016}).  However,
radial migration  
models still vary from one to another on the assumption of initial
conditions, as well as how the bar and spiral arms system are treated.
Still more assumptions are made when a chemical evolution model is
applied to a dynamical one (\citealt{minchev2013} are clear on the
sensitivity of their results to assumptions made in their model).
Adding chemistry to a dynamical model introduces additional uncertainties
that complicate comparison of models to one another and to
observations.  

As a
result, different models can produce conflicting predictions.  For
example, \citet{minchev2013} found that the Sun most likely formed
near somewhere between $R_{\rm GC} = 5.5-7$ kpc, while
\citet{martinez-barbosa2015} found the probability that the Sun formed
inside $R_{\rm GC}= 8$ kpc to be no higher than 30\%.  Furthermore, they
found that when radial migration effects were strong, the Sun was
more likely to migrate from the outer disk, and this only occurred in
specific cases.

In an analysis of FGK stars near the Sun,
\citet{haywood2013} argue for the lack of a detectable influence of
radial migration, at least in the solar neighborhood.  They
argue that the ellipticity of stellar orbits, plus the Sun's nearness
to the transition to the metal-poor outer disk ($R_{\rm GC} \sim 10$ kpc), are
sufficient to explain the presence of both metal-poor and metal-rich
stars in the solar neighborhood.  They furthermore argue that the
Milky Way disk likely formed outside-in, rather than inside-out, based
on the different age distributions of the metal-poor and metal-rich
stars near the Sun.  

Finally, it is still uncertain whether the effect of radial migration
on open clusters is the same as that for field stars.  
Migration has been proposed to explain the presence of
clusters with $[\rm Fe/H] \sim +0.4$ dex near $R_{\rm GC} \sim 7$ kpc, such as NGC
6791; however, a study by \citet{jilkova2012} concluded it was very
unlikely.
On the other hand, \citet{fujii2012} used N-body simulations of
clusters in a Galactic disk and found a timescale of $\sim100$\,Myr 
for the radial migration of open clusters.  In that time
period, they demonstrated a cluster could move $\sim1.5$ kpc from its
birth location, with minimal mass loss.  However, their
simulation did not include the effects of the bar, and so it is not
clear whether their general conclusions apply to clusters inside the
Solar circle, as we consider here.  \citet{martinez-medina2016}
demonstrated that spiral arms can lift open clusters $>1$ kpc above
the Galactic mid-plane, which can explain the tendency of older ($>1$
Gyr) clusters to have large $z$-distances, but they do not explore
variations of cluster $R_{\rm GC}$ in their model.

In conclusion, to fully gauge the impact of radial migration on the
formation and evolution of the disk, the full range of observations of
different stellar populations at different locations in the disk (correcting for
observational biases)  must
be compared with multiple chemo-dynamical models that are built upon
different assumptions and have different uncertainties.  This should
be possible within the next decade, after the many ongoing and
planned spectroscopic surveys of the Galaxy are complete.

\section{Summary}\label{summary}
We have presented an analysis of the inner disk metallicity gradient
based on metallicities of 12 intermediate-aged clusters observed as
part of the {\it Gaia}-ESO Survey.  Our sample triples the number of
clusters inside $R_{\rm GC} = 7$ kpc compared to previous homogeneous
literature studies, and also spans a wider range in age ($\sim1.5$
Gyr) than other studies of objects in this part of the disk (e.g., 
Cepheids in
\citealt{And02inner}). 

The metallicity gradient of our sample, which spans $5.5 < R_{\rm GC} <
8$ kpc, is $-$0.10$\pm$0.02 dex kpc$^{-1}$, consistent within
uncertainties to the gradient measured by APOGEE red giant stars
\citep{hayden2014}.  As our sample stars all lie within 150 pc of the
Galactic plane, the impact of a vertical metallicity gradient on our
results is expected to be less than 0.05 dex, and would only change
the radial gradient by 0.01 dex kpc$^{-1}$.
The metallicity gradient shown by our inner disk sample  is also
consistent with that shown by the larger cluster sample of
\citet{netopil2016} in the range of $6 < R_{\rm GC} < 14$ kpc.  Therefore,
the GES clusters do not support previous claims in the literature that
the metallicity gradient steepens inside the solar circle.  

We have also found that the GES clusters exhibit a trend of increasing
metallicity with cluster age: after correcting for the effect of the
gradient, clusters older than 1 Gyr are 0.06 dex more metal-rich than
clusters aged $<0.5$ Gyr.  Two clusters that exemplify
this effect, Berkeley 44 (age 1.6 Gyr, $[\rm Fe/H] = +0.17\pm0.04$) and
NGC 4815 (age 570 Myr, $[\rm Fe/H] = -0.03\pm0.06$), both reside at
$R_{\rm GC} = 6.9$ kpc, indicating that the dispersion in $[\rm Fe/H]$ with
$R_{\rm GC}$ is $\sim0.2$ dex.

A comparison of the inner disk metallicity gradient to the predictions
of the chemical evolution model of \citet{minchev2014} found
relatively poor agreement (Fig.~4).  The age-metallicity relation for
our sample is consistent with the effects of radial migration  predicted
by \citet{minchev2014}, as also discussed in \citet{netopil2016}, but
the slope and temporal evolution of the gradient predicted by the model is at odds
with the values shown by open clusters in general.  We look forward to a
comparison of cluster and model results when the {\it Gaia}-ESO Survey is complete.   

\begin{acknowledgements}
      The authors thank the referee for comments that improved the
      presentation of this research.
      This work
      was partly supported by the European Union FP7 programme through
      ERC grant number 320360 and by the Leverhulme Trust through
      grant RPG-2012-541. We acknowledge the support from INAF and
      Ministero dell' Istruzione, dell' Universit\`a e della Ricerca
      (MIUR) in the form of the grant ``Premiale VLT 2012'' and ``The Chemical and Dynamical Evolution of the Milky Way and Local Group Galaxies'' (prot. 2010LY5N2T). The results
      presented here benefit from discussions held during the {\it Gaia}-ESO
      workshops and conferences supported by the ESF (European Science
      Foundation) through the GREAT Research Network Programme.
      F.J.E. acknowledges financial support from the ARCHES project 
      (7th Framework of the European Union, n 313146).
      S.V. gratefully acknowledges the support provided by 
      Fondecyt reg. 1130721.  
      U.H. acknowledges support from the Swedish National Space Board (SNSB).
      D.G. gratefully acknowledges support from the Chilean BASAL
      Centro de Excelencia en Astrof\'isica y Tecnolog\'ias Afines
      (CATA) grant PFB-06/2007.
      
\end{acknowledgements}



\clearpage
\begin{table*}
\scriptsize
\caption{Parameters for cluster members.  The complete table is
  available at the CDS.  The first few lines are shown here as a guide
to its contents.} 
\label{member_params}    
\centering
\begin{tabular}{l l l l l l l l l l l }
\hline                 
 & Star & RA (J2000) & DEC (J2000) & $\rm [Fe/H]$ & e($\rm [Fe/H]$) & T$_{\rm eff}$ & e(T$_{\rm eff}$) & log g  & e(log) g & RV\\
Cluster & GES ID & (deg) & (deg)  &  (dex)  & (dex) & (K) & (K) & (dex) & (dex)       &  (km s$^{-1}$) \\
\hline
  Berkeley44 & 19170732+1930555 & 289.2805000 & 19.5154167 & 0.11 & 0.14 & 4998 & 233 & 3.13 & 0.55 & -7.7\\
  Berkeley44 & 19170911+1933256 & 289.2879583 & 19.5571111 & 0.20 & 0.12 & 4943 & 182 & 3.06 & 0.44 & -9.1\\
  Berkeley44 & 19171388+1933333 & 289.3078333 & 19.5592500 & 0.22 & 0.14 & 4996 & 181 & 2.79 & 0.46 & -9.1\\
  Berkeley44 & 19172208+1933254 & 289.3420000 & 19.5570556 & 0.15 & 0.13 & 4880 & 222 & 2.77 & 0.56 & -8.7\\
  Berkeley81 & 19013537-0028186 & 285.3973750 & -0.4718333 & 0.27 & 0.13 & 5167 & 142 & 3.30 & 0.30 & 47.5\\
  Berkeley81 & 19013631-0027447 & 285.4012917 & -0.4624167 & 0.25 & 0.11 & 4991 & 138 & 2.79 & 0.30 & 47.6\\
  Berkeley81 & 19013651-0027021 & 285.4021250 & -0.4505833 & 0.22 & 0.12 & 4939 & 127 & 3.14 & 0.24 & 48.9\\
  Berkeley81 & 19013910-0027114 & 285.4129167 & -0.4531667 & 0.12 & 0.10 & 5012 & 117 & 2.96 & 0.24 & 48.8\\
  Berkeley81 & 19013997-0028213 & 285.4165417 & -0.4725833 & 0.32 & 0.10 & 4970 & 149 & 3.27 & 0.27 & 48.6\\
  Berkeley81 & 19014004-0028129 & 285.4168333 & -0.4702500 & 0.26 & 0.10 & 4940 & 127 & 2.77 & 0.24 & 48.7\\
\hline

\end{tabular}
\end{table*}

\end{document}